\documentclass[%
 aip,
 rsi,
 amsmath,amssymb,
 reprint,%
]{revtex4-2}

\usepackage{graphicx}
\usepackage{upgreek}
\usepackage{siunitx}
\usepackage{dcolumn}
\usepackage{bm}
\usepackage{color}
\usepackage{hyperref}
\hypersetup{
    colorlinks,
    citecolor=blue,    filecolor=blue,
    linkcolor=blue,     urlcolor=blue
}

\begin{document}

\preprint{AIP/123-QED}

\title[]{Symmetric and asymmetric shocked gas jets for laser-plasma experiments}

\author{L. Rovige}
\affiliation{Laboratoire d'Optique Appliqu\'ee, ENSTA, CNRS, Ecole Polytechnique, Institut Polytechnique de Paris, 828 Bv des Maréchaux, 91762 Palaiseau, France.}
\author{J. Huijts}
\affiliation{Laboratoire d'Optique Appliqu\'ee, ENSTA, CNRS, Ecole Polytechnique, Institut Polytechnique de Paris, 828 Bv des Maréchaux, 91762 Palaiseau, France.}
\author{A. Vernier}
\affiliation{Laboratoire d'Optique Appliqu\'ee, ENSTA, CNRS, Ecole Polytechnique, Institut Polytechnique de Paris, 828 Bv des Maréchaux, 91762 Palaiseau, France.}
\author{I. Andriyash}
\affiliation{Laboratoire d'Optique Appliqu\'ee, ENSTA, CNRS, Ecole Polytechnique, Institut Polytechnique de Paris, 828 Bv des Maréchaux, 91762 Palaiseau, France.}
\author{F. Sylla}
\affiliation{SourceLAB, 7 rue de la Croix Martre, 91120, Palaiseau, France.}
\author{V. Tomkus}
\affiliation{Center for Physical Sciences and Technology, Savanoriu Ave. 231, LT-02300, Vilnius, Lithuania.}
\author{V. Girdauskas}
\affiliation{Center for Physical Sciences and Technology, Savanoriu Ave. 231, LT-02300, Vilnius, Lithuania.}
\affiliation{Vytautas Magnus University, K.Donelaicio St. 58. LT-44248, Kaunas, Lithuania.}
 \author{G. Raciukaitis}
 \affiliation{Center for Physical Sciences and Technology, Savanoriu Ave. 231, LT-02300, Vilnius, Lithuania.}
 \author{J. Dudutis}
 \affiliation{Center for Physical Sciences and Technology, Savanoriu Ave. 231, LT-02300, Vilnius, Lithuania.}
 \author{V. Stankevic}
 \affiliation{Center for Physical Sciences and Technology, Savanoriu Ave. 231, LT-02300, Vilnius, Lithuania.}
 \author{P. Gecys}
 \affiliation{Center for Physical Sciences and Technology, Savanoriu Ave. 231, LT-02300, Vilnius, Lithuania.}
\author{J. Faure}
 \email{jerome.faure@ensta-paris.fr}
\affiliation{Laboratoire d'Optique Appliqu\'ee, ENSTA, CNRS, Ecole Polytechnique, Institut Polytechnique de Paris, 828 Bv des Maréchaux, 91762 Palaiseau, France.}

\date{\today}

\begin{abstract}
Shocks in supersonic flows offer both a high-density and sharp density gradients that can be used, for instance, for gradient injection in laser-plasma accelerators. We report on a parametric study of oblique shocks created by inserting a straight axisymmetric section at the end of a supersonic ``de Laval" nozzle. The impact of different parameters such as throat diameter and straight section length is studied through computational fluid dynamics (CFD) simulations. Experimental characterisations of a shocked nozzle are compared to CFD simulations and found to be in good agreement. We then introduce a newly designed asymmetric shocked gas jet, where the straight section is only present on one lateral side of the nozzle, thus providing a gas profile that can be used for density transition injection. In this case, full-3D fluid simulations and experimental measurements are compared and show excellent agreement.
\end{abstract}

\maketitle

\section{Introduction}
The development of laser-plasma accelerators \cite{taji79,esar09} (LPA) requires efforts, not only on the laser-driver part of the system, but also on the shaping of the plasma target. It is indeed necessary to tailor the plasma profile in order to gain more control on injection and acceleration mechanisms and obtain high quality particle beams suitable for applications such as for femtosecond x-ray beams production \cite{rousse04,kneip10,taph12}, particle colliders\cite{schr10}, electron diffraction \cite{he16,faure16} or medical applications \cite{Rigaud2010,lund12}\par
In electron acceleration, the gradient injection scheme relying on a sharp downward density transition \cite{bula98,toma03,suk01,kim04} has been used in numerous experiments and has proven very efficient to increase beam quality and stability. It has been mainly implemented through laser-induced density transition \cite{chien05,faur10} and by inserting a thin blade in the outflow of a supersonic gas jet \cite{schm10,thaury15,swanson17} which results in the formation of a shock-front in the gas profile. Two different regimes can be distinguished according to the relative size between the gradient scale length and the plasma wavelength $\lambda_p$. If $\mathrm{L_{grad}}>\lambda_p$ the injection is due to the reduced wake phase velocity in the density transition region, which facilitates trapping \cite{bula98}. In the case of a sharp transition, $\mathrm{L_{grad}}<\lambda_p$, the plasma wavelength increases abruptly because of the sudden change in plasma density, and some background electrons find themselves trapped in the accelerating phase of the wake\cite{suk01,toma03}. The sharp gradient configuration favours injection in the first bucket which yields shorter electron bunches with narrow energy-spread.\par
Moreover, gas targets are also relevant for ion acceleration experiments in the collisionless shock acceleration \cite{Haberberger2012} regime and magnetic vortex acceleration \cite{naka10} regime which occur in a near-critical plasma. At such high density, the laser beam undergoes a strong absorption and is quickly depleted\cite{sylla13}, therefore these acceleration schemes require a narrow plasma profile with sharp gradients.\par
Shocks in supersonic flows have several advantages making them useful tools to tailor the gas profile in laser-plasma experiments : (i) they can provide high densities with sharp profiles needed in ion acceleration experiments, (ii) this high density can be obtained relatively far from the nozzle which is especially interesting to reduce damage on the target and increase its durability, (iii) they enable the production of gas profiles with a downward density transition followed by a plateau, of particular interest for the gradient injection scheme. \par
As mentioned earlier, most of the experiments relying on the gradient injection method use a blade inserted in the flow after the nozzle. The physics of supersonic gas jets impinged by a blade has been recently thoroughly described \cite{Fan-Chiang20} and such design works well with millimetric-scale targets used in experiments with high-power lasers where the Rayleigh length is relatively long, and thus where distance and positioning constraints are not too stringent. But in high-repetition rate laser-plasma accelerators with an energy of only a few millijoules per pulse, it is necessary to focus the laser tightly in order to achieve relativistic intensities. The targets are therefore scaled down to micrometric dimensions, and the laser is focused at around $\SI{150}{\micro\meter}$ from the nozzle. With such small dimensions, inserting a knife-edge in the flow with good precision can prove difficult. Moreover, as LPA technology advances, questions of stability and reproducibility gain importance in the perspective of applications, and integrating the shock formation in the design of the nozzle would offer a more compact, robust and simple solution than the blade technique.\par
In this paper, we study supersonic shock-nozzles of micrometric-dimensions, relying on the formation of oblique shocks due to the sudden change of flow direction in the final section of the nozzle, with fluid simulations and experimental measurements. A symmetrically shocked design yielding a high on-axis density, with peaked profile\cite{molli16}, is thoroughly studied through simulations, which are validated by an experimental measurement. We then propose a newly designed asymmetrically shocked nozzle intended to provide the density downramp followed by a plateau, necessary to gradient injection. This design is validated through 3D CFD simulations and experimental measurements. We recently showed that this kind of nozzle greatly enhances the long-term stability of a kilohertz laser-plasma accelerators. \cite{rovige20} \par
This paper is organised as follow: in Sec. \ref{sec:theory}, we review some physical principles relevant to the study of supersonic flows and oblique shocks, a simple geometrical model for the on-axis shock position is proposed. Section \ref{sec:methods} presents the methods used for numerical simulations and experimental measurements. Section \ref{sec:sym} is devoted to the study of symmetrically shocked jets, with first a comparison between simulation and measurement, and then a study of the influence of different parameters with simulations. In Sec. \ref{sec:OSS} we present the design of the one-sided shock nozzle, with CFD simulation and an experimental measurement. Finally, Sec. \ref{sec:ccl} summarizes the results and concludes this paper.\par 

\section{\label{sec:theory}Theory}
 
\subsection{1D Isentropic flow}

The design of the gas jet used to produce an oblique shock consists in a converging-diverging \textit{``de Laval"} nozzle in which a straight duct has been added at the exit of the gas jet to abruptly change the direction of the flow. The converging section is attached to a constant pressure reservoir, and the nozzle exhaust leads in a vacuum chamber. This geometry results in a Mach number $M=1$ at the throat, and supersonic flow in the diverging section of the nozzle. The evolution of the flow in such a nozzle has been thoroughly described with a 1D isentropic model \cite{Zucker2002}, and the physics of supersonic nozzles in a context similar to ours has already been studied \cite{semu01,Schmid2012}, therefore, we will limit ourselves to recalling the main results of the isentropic expansion model. The flow parameters, namely the temperature $T$, the pressure $p$ and the density $\rho$, can be expressed according to the Mach number M and their initial value in the reservoir\cite{Zucker2002}. The coefficient $\gamma$ is the specific heat ratio of the gas, which is 5/3 for monoatomic gases and 7/5 for diatomic gases. $A_t$ is the cross-section area of the nozzle at the throat, and $A$ the area at the interest point. \par

\begin{gather}
\label{eqn:surfmach}
\frac{A_t}{A}=M\left[1+\frac{\gamma-1}{\gamma+1}(M^2-1)\right]^{-\frac{\gamma+1}{2(\gamma-1)}}\\
\frac{T}{T_0}=\left(1+\frac{\gamma-1}{2}M^2 \right)^{-1}\\
\frac{p}{p_0}=\left(1+\frac{\gamma-1}{2}M^2 \right)^{-\frac{\gamma}{\gamma-1}}\\
\label{eqn:densmach}
\frac{\rho}{\rho_0}=\left(1+\frac{\gamma-1}{2}M^2 \right)^{-\frac{1}{\gamma-1}}
\end{gather}
It appears that all the physical quantities are determined by the ratio between the area of the nozzle at the throat and the area at the considered point. Equation \ref{eqn:surfmach} will be of particular interest for our study as it allows us to determine the Mach number which is one of the governing factor of the behavior of oblique shocks. Moreover, equation \ref{eqn:densmach} shows that the density decreases as the nozzle section (and therefore the Mach number) increases. In a simple supersonic nozzle, the same behavior happens at the exit, when the flow expands freely into vacuum, leading to a density which decreases quickly with the distance z, and a degraded profile. The use of oblique shocks, as described in the next section, makes it possible to compensate for this expansion in order to obtain high densities further from the nozzle.\par
It is important to note that this model does not take into account the effects of the boundary layer, i.e the region near the wall where the flow velocity transitions from 0\% to 90\% of the center velocity and where the isentropic assumption is not valid.

\subsection{\label{sec:geomod}Oblique shock theory and geometric model of on-axis peak density position}
\begin{figure}[th!]
     \centering
            \includegraphics[width=0.98\linewidth]{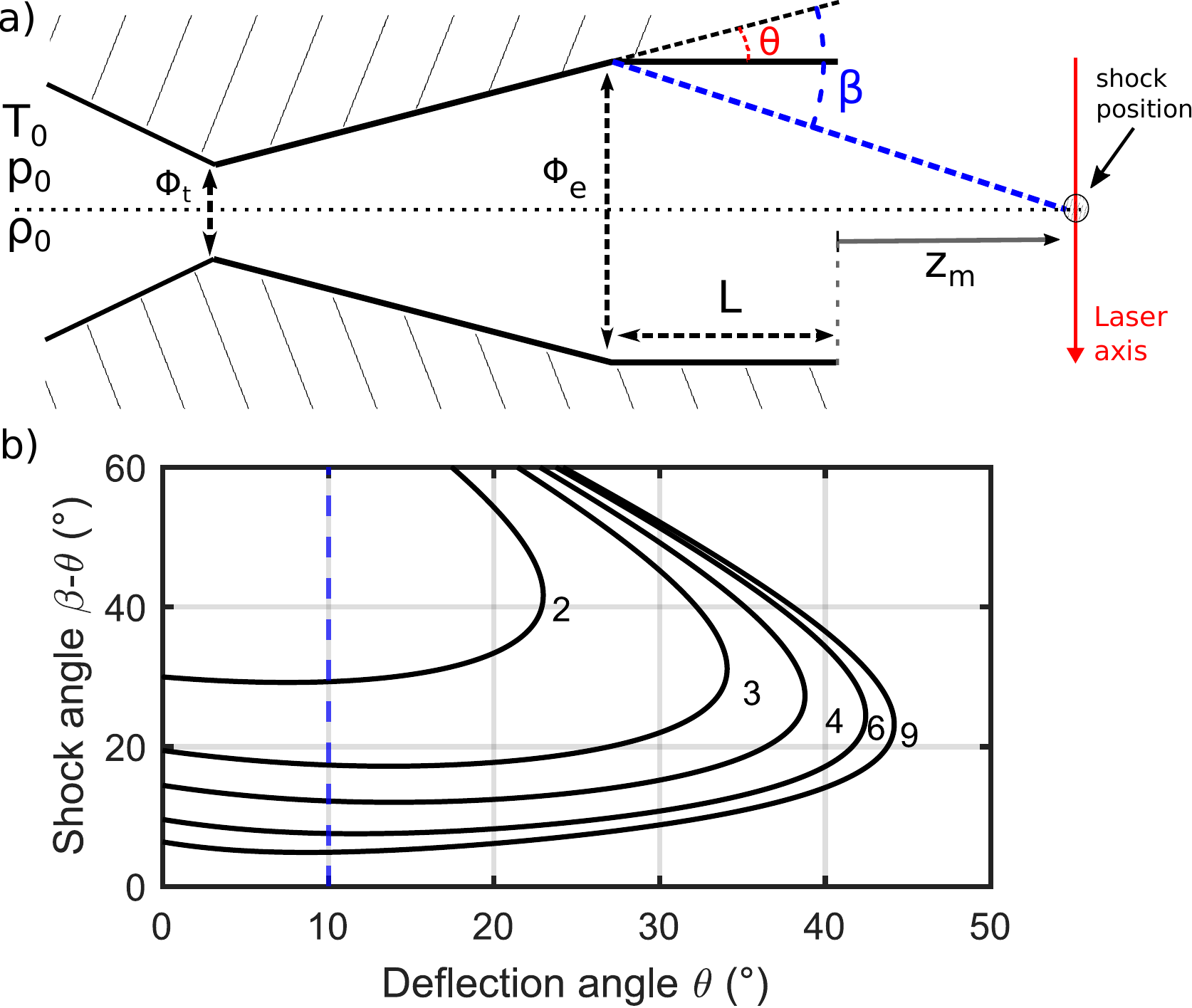}
        \caption{\label{fig:thang} a) Schematic description of an oblique shock formation in a shock nozzle   b) Shock angle as a function of the deflection angle for different Mach numbers. The dashed blue line represents the angle of $10^\circ$ used later in the design of our jets.}   
\end{figure}
A shock in a supersonic flow is characterized by a sudden reduction of the Mach number at a certain position, leading to the compression of the gas in the shocked region. This compression leads to higher density which is of interest for gas target design. When a supersonic flow changes direction abruptly, such as when encountering a wedge with a moderate (we will see later what is moderate in this case) deflection angle $\theta$, it generates an oblique shock-wave originating from the corner of the wedge and at an angle $\beta$ to the original flow direction. We propose to study the configuration sketched in Fig. \ref{fig:thang}a where a straight duct added at the end of the diverging section of a ``de Laval" nozzle induces a shock-front of angle $\beta-\theta$ with the longitudinal axis. The shock-fronts then converge on-axis at a distance $z_m$ from the nozzle exit determined by the shock angle and the length of the straight duct. This configuration yields peaked gas profile with high density relatively far from the nozzle. The relation between the shock angle $\beta$,  the deflection angle $\theta$ and the Mach number before the shock $M_1$ is given by equation \ref{eqn:beta}\cite{Zucker2002,Liepmann2013} :\par
\begin{equation}
\mathrm{tan\,\theta = 2\,cot\,\beta\,\frac{M_1^2\,sin^2\,\beta-1}{M_1^2\,(\gamma+cos\,2\beta)+2)}}
    \label{eqn:beta}
\end{equation}
Equation \ref{eqn:beta} does not allow to explicitly express $\beta$ according to $\theta$ and $\mathrm{M_1}$, but we can determine it graphically. The solution of $\beta-\theta$ according to $\theta$ for different Mach numbers is displayed in Fig.~\ref{fig:thang}b. For each deflection angle there are two solutions, one with a low shock angle, corresponding to the \textit{weak shock} solution leading to a still supersonic Mach number after the shock $\mathrm{M_2}>1$, and one with a higher shock angle, corresponding to the \textit{strong shock} case, with a subsonic downstream flow. Even if no clear mathematical criterion is known, in practice, the \textit{weak shock} case is almost always observed in experiments, as the \textit{strong shock} requires a higher pressure downstream \cite{courant48} obtained only in specific conditions. In our case where a supersonic flow expands into near-vacuum, the weak shock will therefore occur. Equation \ref{eqn:beta} does not have any solution for deflection angles $\theta>\theta_{max}$ depending on the Mach number, in this case the shock solution is not an oblique shock but a detached bow-shock\cite{Zucker2002}.\par
It is then possible to determine geometrically the on-axis position of the shock, thanks to the angle $\beta-\theta$ :
\begin{equation}
z_m=\mathrm{\frac{\phi_e/2}{\tan{(\beta-\theta)}}-L}
    \label{eqn:zm}
\end{equation}
Where L is the length of the straight section at the end of the diverging section, and $\phi_e$ is the exit diameter of the nozzle (see Fig. \ref{fig:thang}a). Even though the oblique shock originates from the corner of the wedge, the on-axis shock position $z_m$ is given with respect to the exit of the nozzle, (hence the subtraction of L) because this is the relevant quantity from an experimental point of view.\par
In order to have a shock position far from the nozzle and preserve its integrity, the shock angle should be kept small. As is clear from \ref{fig:thang}b, this can be obtained through a sufficiently high Mach number ($>3$) at the end of the diverging section (determined by $A_e/A_t$). Although very useful to determine the above principles, this geometrical model does not give indications on the density obtained, nor on the effect of the length of the straight section. Numerical simulations are therefore needed to understand these characteristics. \par

\section{\label{sec:methods}Methods}
\begin{figure*}[t!]
     \centering
         \includegraphics[width=0.98\linewidth]{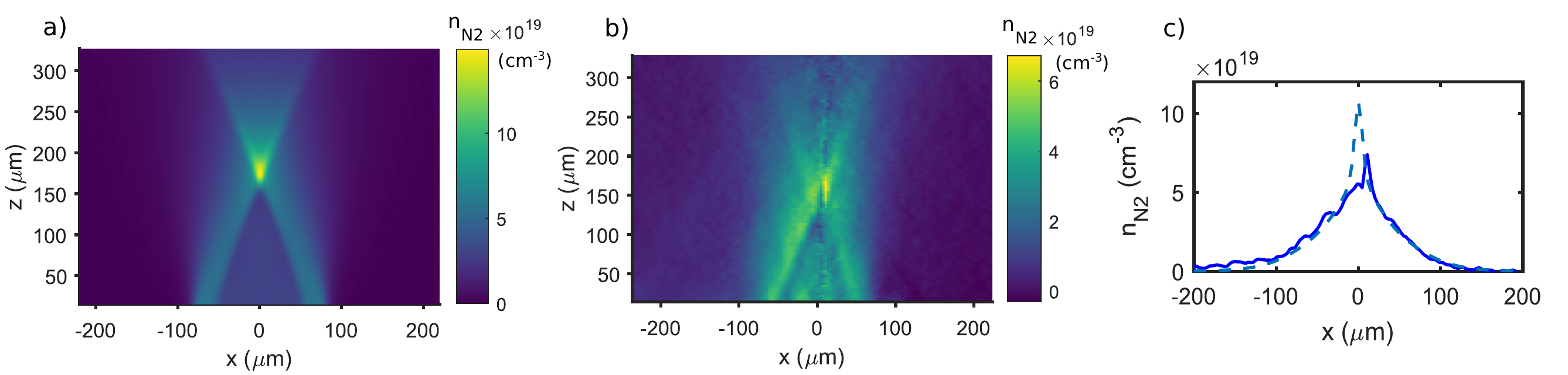}
        \caption{\label{fig:comp} a) Simulated and   b) experimental nitrogen molecular density map of a symmetric shock nozzle with a backing pressure $P_{back}=50$\,bar   c) Comparison of the simulated (dashed) and measured (solid) density profiles at $z=z_m$}   
\end{figure*}
The simulations are carried out with the CFD software ANSYS Fluent which solves the Navier-Stokes equations. The k-$\upomega$ shear stress transport (k-$\upomega$ SST) turbulence model\cite{wilcox98,menter93} is used. It is a robust and efficient model which uses the k-$\upomega$ formulation near the boundary layers, and switches to the k-$\upepsilon$ formulation in the free-stream. Simulations are performed using nitrogen $N_2$. Both 2D-axisymmetric and 3D geometry are used depending on the symmetries of the design. The mesh is refined around regions of interest, and is composed of $\sim 10^5$ cells in the 2D cases and $\sim 5\times10^6$ cells for the 3D simulations. A convergence study has been performed to ensure that further refining of the mesh does not significantly change the solution. Full-multigrid initialization is used to obtain an initial guess of the solution thus allowing faster convergence. \par
The experimental characterization of symmetric nozzles is performed in gaseous nitrogen, because this is the primary gas used in our experiments, and its high refractive index provides a high signal level. The 2D-phase map of the gas jet is measured with a quadriwave lateral shearing interferometer\cite{primot1995,Bohle2014} (QWLSI), the density map is then retrieved  via Abel inversion of the measured phase map. This method cannot be applied to non-axisymetric jets, because the Abel inversion algorithm requires cylindrical symmetry. Instead the measurement is performed by ionizing the gas with a laser: the created plasma column has (upon approximations) an axis of symmetry along the laser beam propagation axis.  To do so, a laser pulse is sent into the gas jet, at the desired probing distance. A 25\,fs, 2\,mJ pulse is focused to a $\SI{6}{\micro\meter}$ FWHM spot, therefore reaching an intensity of $\sim \SI{1.3e17}{\watt\per\square\centi\meter}$ which is one order of magnitude higher than the intensity necessary to ionize nitrogen into $N^{5+}$. The plasma produced by the beam is illuminated from the side by a probe beam, and imaged onto the QWLSI. The plasma density profile can then be derived from the phase maps via Abel inversion, assuming radial symmetry around the laser-axis. This is possible under the assumption that the evolution of the  gas density is low over the plasma column radial dimension. Still, the angle between the oblique shock and the normal to the laser propagation axis induces a slight asymmetry in the plasma channel that we neglect.\par

\section{\label{sec:sym}Symmetric shocked jets}

When a straight section is added at the end of a ``de Laval" nozzle, as pictured in Fig. \ref{fig:thang}a, oblique shocks arise from the whole outer diameter of the jet, and converge to a point on the axis, resulting in a very dense and narrow gas profile. The study of symmetric shock-jets can be performed in 2D-axisymmetric geometry. This understanding can then be used in the context of the asymmetric shock-jet of section \ref{sec:OSS}, which require full-3D simulations.\par

\subsection{Comparison between measurement and simulation}

In order to validate our CFD simulations, we have performed measurements of the gas density profile of a symmetric shock-jet. Figure \ref{fig:comp} shows the results of the measurement performed on a jet with  $\phi_t=\SI{60}{\micro\meter}$, $\phi_e=\SI{180}{\micro\meter}$, and a $10^\circ$ diverging section, with a straight duct length $\mathrm{L}=\SI{100}{\micro\meter}$, and the comparison with the simulated profile. The isentropic model predicts a Mach number of 3.8 at the end of the diverging section, which would result in a $13^\circ \, \beta-\theta$ shock angle. The geometric model of section \ref{sec:geomod} predicts an on-axis shock position at $z_{m,th}=289\,\mathrm{\mu m}$.\par
The measurement indeed shows the convergence of shock structures on the jet axis, yielding a substantially high density and peaked profile. The simulation prediction of the position of the shock is   $\mathrm{z_{m,s}}=\SI{176}{\micro\meter}$ while the measured position is $\mathrm{z_{m,m}}=\SI{166}{\micro\meter}$, which shows a fairly good agreement. These values are significantly lower than predicted by the geometrical model, indicating that the boundary layer plays an important role in the physics of micrometric jets. In the simulation, the center Mach number at the end of the diverging section is 3.6, and the flow velocity decreases near the walls. The simulated and measured gas density transverse profiles at the on-axis shock position are showed on Fig. \ref{fig:comp}.c. Both profiles have similar widths, but in the experimental case, the peak density is significantly lower. This could be due to an insufficient resolution (phase resolution is $\SI{3.2}{\micro\meter}$) combined with the high on-axis noise of the Abel inversion used to retrieve the density from the measured phase. Still, the good overall agreement between measurement and simulation validates the use of CFD simulations for the design and study of shocked gas jets.\par

\subsection{Parametric study}

We numerically study the influence of two parameters, the length of the final straight duct L, and the diameter of the throat $\phi_t$, on the position $z_{max}$ where the shock structures meet on the axis thus forming a peaked density profile, and on the density $n_{max}$ at this position. The exit diameter is fixed at $\SI{300}{\micro\meter}$, the angle of the diverging section is fixed at 10$^\circ$, the origin of the z axis is the exit of the nozzle\par
\begin{figure}[th!]
     \centering
         \includegraphics[width=0.90\linewidth]{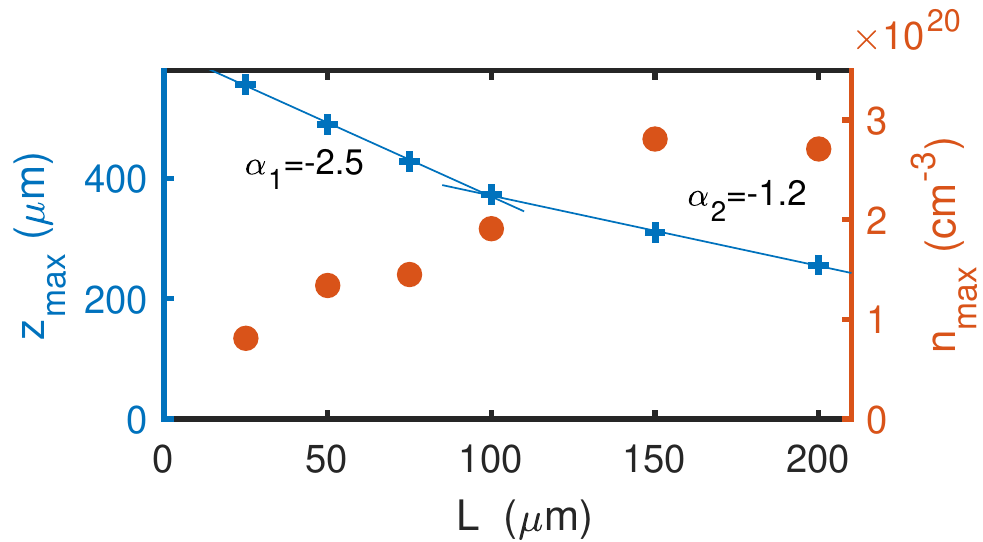}
        \caption{\label{fig:flatduct}Simulated on-axis shock position (blue cross) and linear fit for the two regimes, nitrogen molecular density at this position (orange dots), as a function of the length of the final straight duct L. $\alpha_1$ and $\alpha_2$ are the slopes of the linear fits. Throat diameter is fixed at $\mathrm{\phi_t}=100\,\mathrm{\upmu m}$. Simulations are performed in nitrogen with a backing pressure $P_{back}=50$\,bar.}   
\end{figure}

A numerical study of the effect of the straight duct length, in Fig. \ref{fig:flatduct} is of particular interest, as no information on the matter is given by the theoretical model. In Fig.\,\ref{fig:flatduct}, it appears that for L\,$<\SI{100}{\micro\meter}$ an increase in the length of the straight section leads to the shock being formed closer to the nozzle, with a slope of -2.5. For higher values of L, a further increase of the straight duct length has almost no significant effect on the position of the shock other than the nozzle's exit being brought closer to it due to the length increase. On the other hand, the maximum density increases with L, until it saturates at L=$\SI{150}{\micro\meter}$. These results show that a compromise on the final duct length has to be made to obtain high density sufficiently far away from the nozzle to prevent from damaging. In our configuration, values of L larger than $\SI{150}{\micro\meter}$ do not provide any benefit. \par
The influence of L on the shock can be explained by the fact that in the nozzle, the flow direction is not homogeneous. On the center of the nozzle, the gas flows parallel to the axis, while near the walls the flow lines have a $10^{\circ}$ angle corresponding to the expansion angle of the nozzle. In the case where L is very short, only the outer flow lines will contribute to the shock, because the inner ones will not ``see" the change in direction. And if L is increased, more flow lines will coalesce into the shock front, which will therefore be stronger. Moreover, the effective deflection angle for these supplementary flow lines is smaller, which results in a larger shock angle (see Fig. \ref{fig:thang}) which could explain the decrease of $\mathrm{z_m}$ with L observed in Fig.\,\ref{fig:flatduct}.
\par

\begin{figure}[th!]
     \centering
         \includegraphics[width=0.95\linewidth]{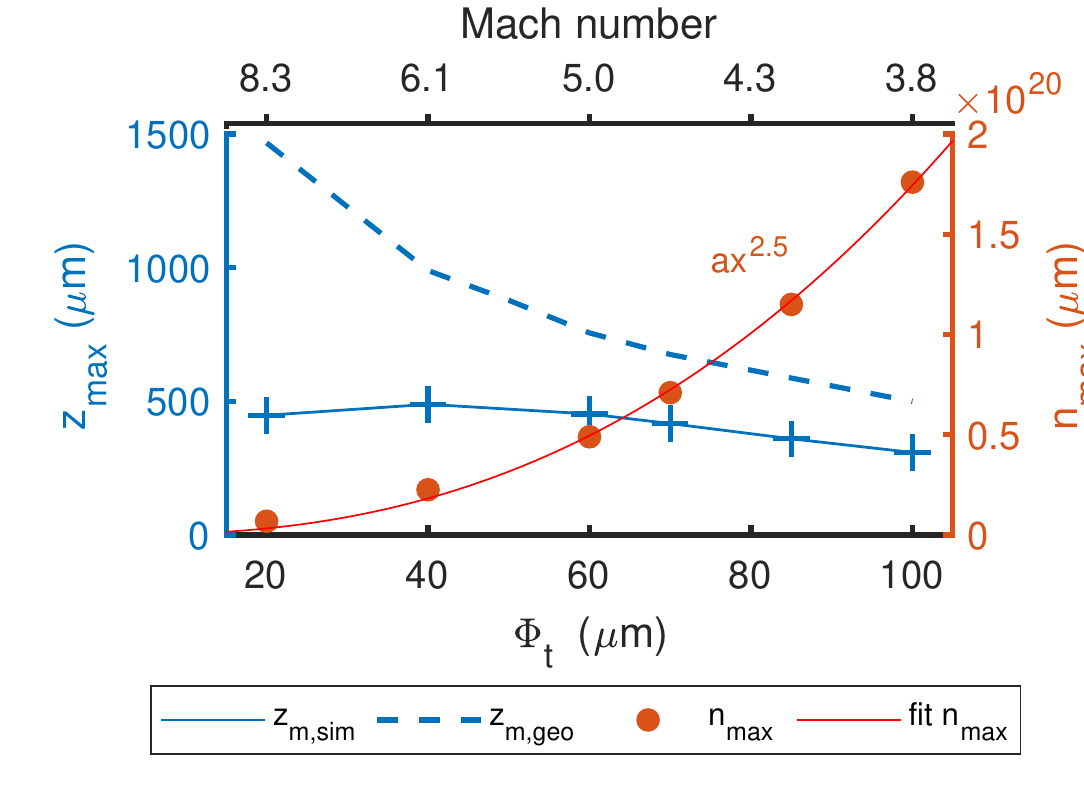}
        \caption{\label{fig:throat} Evolution of the on-axis shock position (blue cross) and of the maximum density at this point (orange dot) as a function of the throat diameter, and corresponding Mach number at the end of the diverging section. The blue dashed line represent the predictions of the geometrical model of section \ref{sec:theory}. The orange line is a power fit of the maximum density data. Simulations are performed in nitrogen with a backing pressure $P_{back}=50$\,bar} 
\end{figure}

Figure \ref{fig:throat} shows the numerical evolution of those two same quantities, shock position and maximum density, as well as the prediction of the geometric model of section \ref{sec:geomod} for the shock position, as a function of the throat diameter, with the same geometry as before and a fixed value of $\mathrm{L}=\SI{100}{\micro\meter}$. Reducing the throat diameter while keeping the same exit diameter leads to an increase of the Mach number, as can be deduced from Eq. \ref{eqn:surfmach}, which can be interesting in order to increase the distance of the density peak $z_m$. It appears that the simple geometric model correctly predicts the tendency, despite an offset, of an increase in the shock position when the throat diameter $\phi_t$ decreases, for diameters larger than $\SI{60}{\micro\meter}$. For smaller $\phi_t$ the flow is governed by boundary layers, which are not considered in the simple model, and the shock position saturates around $z_{max}=\SI{500}{\micro\meter}$ and even decreases for the smallest diameter considered. Moreover, the offset of the geometric model compared to the simulations for the higher $\phi_t$ values can be explained again by the effect of the boundary layer, which induces a lower Mach number than calculated with the 1D-isentropic model in the region near the walls, therefore increasing the shock angle.\par
The maximum density increases with the throat diameter, but this process is largely governed by the evident rise of mass flow rate at the throat due to the larger cross section.\par 
This parametric study shows that by modifying the length of the straight section and the throat diameter, it is possible to control the peak density and its distance from the nozzle. But both nozzle's features have opposite impact on the flow characteristics, therefore a compromise corresponding to the experimental requirement has to be found. With a backing pressure $P_{back}=50$\,bar, nitrogen density up to \SI{2.8e20}{\per\cubic\centi\meter}  at $z_m=310\,\upmu\mathrm{m}$ is predicted with this design, which corresponds to a plasma density $n_e=\SI{2.8e21}{\per\cubic\centi\meter}=1.6\,n_c$ at $\lambda_0=800\,\mathrm{nm}$ after ionization of $N_2$ into $N^{5+}$. Symmetric shock nozzles therefore make it possible to reach near-critical to over-critical densities without the need to use a high-pressure compressor. Moreover, with a 150\,bar backing pressure, which can be obtained directly at the exhaust of commercial gas bottles, a density even three times higher would be achievable. \par

\begin{figure*}[th!]
     \centering
         \includegraphics[width=0.97\linewidth]{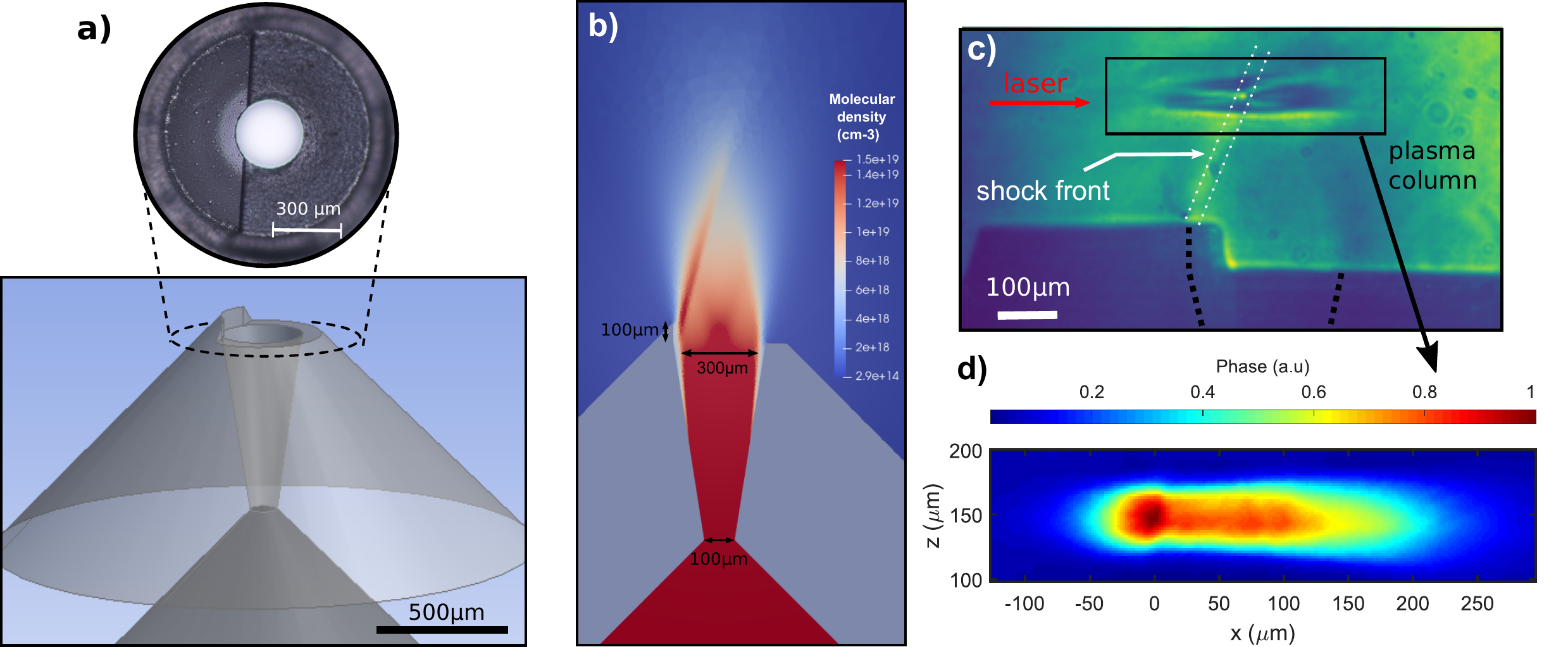}
        \caption{\label{fig:OSS}a) 3D-model of a one-sided shock nozzle, with a zoom on a top-view picture of the nozzle taken with an optical microscope.   b) Slice of the nitrogen density map from 3D CFD Fluent simulation, with a backing pressure of 15\,bar.	c) Experimental shadowgraphic image of the plasma. The black dotted line suggests the inner walls of the nozzle, the white dotted lines highlight the shock front.		d) Normalized phase map of the plasma channel obtained by quadriwave lateral shearing interferometry at $z=\SI{150}{\micro\meter}$ from the nozzle's exit.}   
\end{figure*}

\section{\label{sec:OSS}One-Sided Shocked Jets}

In this section, we present a design using an oblique shock only on one side of the nozzle in order to tailor the gas profile for injection in the sharp density downward transition induced by the shock structure. This design is asymmetric, and therefore 2D-axisymetric simulations can no longer be used. It is necessary to perform more extensive full-3D CFD simulations.\par
\begin{figure}[th!]
     \centering
         \includegraphics[width=1\linewidth]{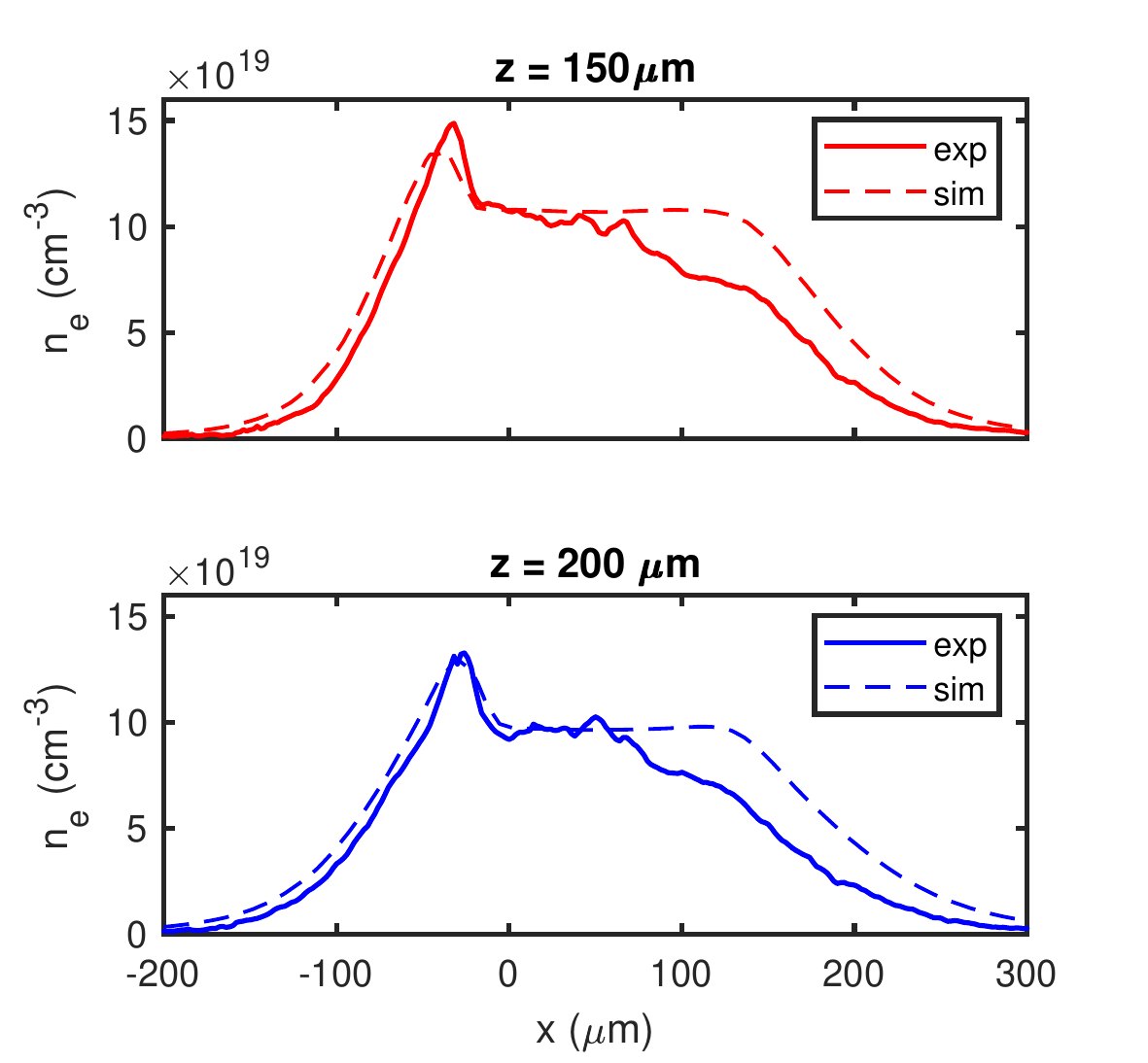}
        \caption{\label{fig:OSS_sim_exp} Comparison of measured and simulated plasma profile obtained with a one-sided shock nozzle using nitrogen with a backing pressure of 15\,bar at a distance of a) $\SI{150}{\micro\meter}$  and b)  $\SI{200}{\micro\meter}$  from the nozzle's exit} 
\end{figure}

The manufacture of such small nozzles with asymmetric features has been made possible by the use of the femtosecond laser-assisted selective etching (FLSE) technique \cite{Marcinkevicius2001,Tomkus18}.   Figure \ref{fig:OSS}b shows the simulated density map obtained by using nitrogen with a backing pressure of 15\,bar. The straight section here shown on the left side was designed to generate an additional shock propagating at an angle with respect to the jet axis. In the simulation, the shock angle is $\beta-\theta \sim 14^\circ$ which is in good agreement with the theory presented in Sec.\,\ref{sec:theory} that predicts an angle of $13^\circ$. The slight difference can be explained by the effects of boundary layers that are not taken into account by the 1D-isentropic model. Figure \ref{fig:OSS_sim_exp} compares the density profile obtained in the simulations with the one retrieved from the measured phase maps in a nitrogen plasma, at two different distances from the nozzle's exit. Fluid simulations give us the $N_2$ molecular density, from which we retrieve the corresponding plasma density by assuming ionization up to $N^{5+}$. The simulation shows a very good agreement with the measured profile as well as with the absolute density value. At $z=\SI{150}{\micro\meter}$ the measured length of the density downward transition is $\SI{16}{\micro\meter}$ ($\SI{18}{\micro\meter}$ in the simulation) for a density drop of 26\% (21\% in the simulation). At $z=\SI{200}{\micro\meter}$ the measured length of the density downward transition is $\SI{26}{\micro\meter}$ ($\SI{27}{\micro\meter}$ in the simulation) for a density drop of 31\% (24\% in the simulation). This typical shock length corresponds to only a few plasma wavelengths in our high density regime ($\uplambda_p\sim\SI{3}{\micro\meter}$ at $n_e=\SI{1.4e20}{\per\cubic\centi\meter}$) which is well suited to density gradient injection. It also appears that after $x=\SI{75}{\micro\meter}$ there is a decrease in the measured density that is not predicted by the simulation. It has been verified that this is not due to a decrease in intensity by scanning the relative position of the jet with respect to the laser focus. This could be explained by different factors such as defects in the inside geometry of the nozzle or a slight angle between the laser direction and the normal to the shock structure. \par

\section{\label{sec:ccl}Conclusion}
We have presented a CFD-parametric study of the effect of different parameters on the behaviour of oblique shock created by a straight section at the end of a supersonic nozzle. Through the modification of the straight duct length and throat diameter, it is possible to control the position and maximum density of the shocked region. Reducing the throat diameter leads to an increase of the Mach number, a lower shock angle and moves the focus further away from the nozzle, at the price of a reduced mass flow and therefore smaller density. Reducing the straight duct length will decrease the shock angle, but will also decrease the shock strength. \par
We then presented a new design of shocked gas jet, with an oblique shock on only one side, therefore providing a downward density gradient at the beginning of a transverse path in the flow, that can be used for the gradient injection scheme. We validated this design through 3D CFD simulations, and with experimental characterisation of a $\SI{300}{\micro\meter}$ diameter one-sided shock jet. The knowledge about the behaviour of oblique shock obtained through the 2D-axisymetric simulations of Sec.\,\ref{sec:sym} can be applied to the one-sided shock case, and provides us with the general laws to modify the characteristics of the density gradient. This new asymmetric design is particularly well suited to small targets, where inserting a knife-edge in the flow can be difficult. Because it provides a single-piece solution for shock-formation, its robustness and ease-of-implementation can benefit a large number of configurations in laser-plasma experiments. 

\begin{acknowledgments}

We acknowledge  Laserlab-Europe, H2020 EC-GA 654148 and the Lithuanian Research Council under grant agreement No. S-MIP-17-79.
\end{acknowledgments}

\section*{Data Availability}
The data that support the findings of this study are available from the corresponding author upon reasonable request.

\providecommand{\noopsort}[1]{}\providecommand{\singleletter}[1]{#1}%


\begin{thebibliography}{36}%
\makeatletter
\providecommand \@ifxundefined [1]{%
 \@ifx{#1\undefined}
}%
\providecommand \@ifnum [1]{%
 \ifnum #1\expandafter \@firstoftwo
 \else \expandafter \@secondoftwo
 \fi
}%
\providecommand \@ifx [1]{%
 \ifx #1\expandafter \@firstoftwo
 \else \expandafter \@secondoftwo
 \fi
}%
\providecommand \natexlab [1]{#1}%
\providecommand \enquote  [1]{``#1''}%
\providecommand \bibnamefont  [1]{#1}%
\providecommand \bibfnamefont [1]{#1}%
\providecommand \citenamefont [1]{#1}%
\providecommand \href@noop [0]{\@secondoftwo}%
\providecommand \href [0]{\begingroup \@sanitize@url \@href}%
\providecommand \@href[1]{\@@startlink{#1}\@@href}%
\providecommand \@@href[1]{\endgroup#1\@@endlink}%
\providecommand \@sanitize@url [0]{\catcode `\\12\catcode `\$12\catcode
  `\&12\catcode `\#12\catcode `\^12\catcode `\_12\catcode `\%12\relax}%
\providecommand \@@startlink[1]{}%
\providecommand \@@endlink[0]{}%
\providecommand \url  [0]{\begingroup\@sanitize@url \@url }%
\providecommand \@url [1]{\endgroup\@href {#1}{\urlprefix }}%
\providecommand \urlprefix  [0]{URL }%
\providecommand \Eprint [0]{\href }%
\providecommand \doibase [0]{https://doi.org/}%
\providecommand \selectlanguage [0]{\@gobble}%
\providecommand \bibinfo  [0]{\@secondoftwo}%
\providecommand \bibfield  [0]{\@secondoftwo}%
\providecommand \translation [1]{[#1]}%
\providecommand \BibitemOpen [0]{}%
\providecommand \bibitemStop [0]{}%
\providecommand \bibitemNoStop [0]{.\EOS\space}%
\providecommand \EOS [0]{\spacefactor3000\relax}%
\providecommand \BibitemShut  [1]{\csname bibitem#1\endcsname}%
\let\auto@bib@innerbib\@empty
\bibitem [{\citenamefont {Tajima}\ and\ \citenamefont {Dawson}(1979)}]{taji79}%
  \BibitemOpen
  \bibfield  {author} {\bibinfo {author} {\bibfnamefont {T.}~\bibnamefont
  {Tajima}}\ and\ \bibinfo {author} {\bibfnamefont {J.~M.}\ \bibnamefont
  {Dawson}},\ }\bibfield  {title} {\enquote {\bibinfo {title} {Laser electron
  accelerator},}\ }\href {https://doi.org/10.1103/PhysRevLett.43.267}
  {\bibfield  {journal} {\bibinfo  {journal} {Phys. Rev. Lett.}\ }\textbf
  {\bibinfo {volume} {43}},\ \bibinfo {pages} {267--270} (\bibinfo {year}
  {1979})}\BibitemShut {NoStop}%
\bibitem [{\citenamefont {Esarey}, \citenamefont {Schroeder},\ and\
  \citenamefont {Leemans}(2009)}]{esar09}%
  \BibitemOpen
  \bibfield  {author} {\bibinfo {author} {\bibfnamefont {E.}~\bibnamefont
  {Esarey}}, \bibinfo {author} {\bibfnamefont {C.~B.}\ \bibnamefont
  {Schroeder}},\ and\ \bibinfo {author} {\bibfnamefont {W.~P.}\ \bibnamefont
  {Leemans}},\ }\bibfield  {title} {\enquote {\bibinfo {title} {Physics of
  laser-driven plasma-based electron accelerators},}\ }\href
  {https://doi.org/10.1103/RevModPhys.81.1229} {\bibfield  {journal} {\bibinfo
  {journal} {Rev. Mod. Phys.}\ }\textbf {\bibinfo {volume} {81}},\ \bibinfo
  {pages} {1229--1285} (\bibinfo {year} {2009})}\BibitemShut {NoStop}%
\bibitem [{\citenamefont {Rousse}\ \emph {et~al.}(2004)\citenamefont {Rousse},
  \citenamefont {Phuoc}, \citenamefont {Shah}, \citenamefont {Pukhov},
  \citenamefont {Lefebvre}, \citenamefont {Malka}, \citenamefont {Kiselev},
  \citenamefont {Burgy}, \citenamefont {Rousseau}, \citenamefont {Umstadter},\
  and\ \citenamefont {Hulin}}]{rousse04}%
  \BibitemOpen
  \bibfield  {author} {\bibinfo {author} {\bibfnamefont {A.}~\bibnamefont
  {Rousse}}, \bibinfo {author} {\bibfnamefont {K.~T.}\ \bibnamefont {Phuoc}},
  \bibinfo {author} {\bibfnamefont {R.}~\bibnamefont {Shah}}, \bibinfo {author}
  {\bibfnamefont {A.}~\bibnamefont {Pukhov}}, \bibinfo {author} {\bibfnamefont
  {E.}~\bibnamefont {Lefebvre}}, \bibinfo {author} {\bibfnamefont
  {V.}~\bibnamefont {Malka}}, \bibinfo {author} {\bibfnamefont
  {S.}~\bibnamefont {Kiselev}}, \bibinfo {author} {\bibfnamefont
  {F.}~\bibnamefont {Burgy}}, \bibinfo {author} {\bibfnamefont {J.-P.}\
  \bibnamefont {Rousseau}}, \bibinfo {author} {\bibfnamefont {D.}~\bibnamefont
  {Umstadter}},\ and\ \bibinfo {author} {\bibfnamefont {D.}~\bibnamefont
  {Hulin}},\ }\bibfield  {title} {\enquote {\bibinfo {title} {Production of a
  kev x-ray beam from synchrotron radiation in relativistic laser-plasma
  interaction},}\ }\href {https://doi.org/10.1103/PhysRevLett.93.135005}
  {\bibfield  {journal} {\bibinfo  {journal} {Phys. Rev. Lett.}\ }\textbf
  {\bibinfo {volume} {93}},\ \bibinfo {pages} {135005} (\bibinfo {year}
  {2004})}\BibitemShut {NoStop}%
\bibitem [{\citenamefont {Kneip}\ \emph {et~al.}(2010)\citenamefont {Kneip},
  \citenamefont {McGuffey}, \citenamefont {Martins}, \citenamefont {Martins},
  \citenamefont {Bellei}, \citenamefont {Chvykov}, \citenamefont {Dollar},
  \citenamefont {Fonseca}, \citenamefont {Huntington}, \citenamefont
  {Kalintchenko}, \citenamefont {Maksimchuk}, \citenamefont {Mangles},
  \citenamefont {Matsuoka}, \citenamefont {Nagel}, \citenamefont {Palmer},
  \citenamefont {Schreiber}, \citenamefont {{Ta Phuoc}}, \citenamefont
  {Thomas}, \citenamefont {Yanovsky}, \citenamefont {Silva}, \citenamefont
  {Krushelnick},\ and\ \citenamefont {Najmudin}}]{kneip10}%
  \BibitemOpen
  \bibfield  {author} {\bibinfo {author} {\bibfnamefont {S.}~\bibnamefont
  {Kneip}}, \bibinfo {author} {\bibfnamefont {C.}~\bibnamefont {McGuffey}},
  \bibinfo {author} {\bibfnamefont {J.~L.}\ \bibnamefont {Martins}}, \bibinfo
  {author} {\bibfnamefont {S.~F.}\ \bibnamefont {Martins}}, \bibinfo {author}
  {\bibfnamefont {C.}~\bibnamefont {Bellei}}, \bibinfo {author} {\bibfnamefont
  {V.}~\bibnamefont {Chvykov}}, \bibinfo {author} {\bibfnamefont
  {F.}~\bibnamefont {Dollar}}, \bibinfo {author} {\bibfnamefont
  {R.}~\bibnamefont {Fonseca}}, \bibinfo {author} {\bibfnamefont
  {C.}~\bibnamefont {Huntington}}, \bibinfo {author} {\bibfnamefont
  {G.}~\bibnamefont {Kalintchenko}}, \bibinfo {author} {\bibfnamefont
  {A.}~\bibnamefont {Maksimchuk}}, \bibinfo {author} {\bibfnamefont {S.~P.~D.}\
  \bibnamefont {Mangles}}, \bibinfo {author} {\bibfnamefont {T.}~\bibnamefont
  {Matsuoka}}, \bibinfo {author} {\bibfnamefont {S.~R.}\ \bibnamefont {Nagel}},
  \bibinfo {author} {\bibfnamefont {C.~A.~J.}\ \bibnamefont {Palmer}}, \bibinfo
  {author} {\bibfnamefont {J.}~\bibnamefont {Schreiber}}, \bibinfo {author}
  {\bibfnamefont {K.}~\bibnamefont {{Ta Phuoc}}}, \bibinfo {author}
  {\bibfnamefont {A.~G.~R.}\ \bibnamefont {Thomas}}, \bibinfo {author}
  {\bibfnamefont {V.}~\bibnamefont {Yanovsky}}, \bibinfo {author}
  {\bibfnamefont {L.~O.}\ \bibnamefont {Silva}}, \bibinfo {author}
  {\bibfnamefont {K.}~\bibnamefont {Krushelnick}},\ and\ \bibinfo {author}
  {\bibfnamefont {Z.}~\bibnamefont {Najmudin}},\ }\href
  {https://doi.org/10.1038/nphys1789} {\bibfield  {journal} {\bibinfo
  {journal} {Nat. Phys.}\ }\textbf {\bibinfo {volume} {6}},\ \bibinfo {pages}
  {980--983} (\bibinfo {year} {2010})}\BibitemShut {NoStop}%
\bibitem [{\citenamefont {{Ta Phuoc}}\ \emph {et~al.}(2012)\citenamefont {{Ta
  Phuoc}}, \citenamefont {Corde}, \citenamefont {Thaury}, \citenamefont
  {Malka}, \citenamefont {Tafzi}, \citenamefont {Goddet}, \citenamefont
  {C.Shah}, \citenamefont {Sebban},\ and\ \citenamefont {Rousse}}]{taph12}%
  \BibitemOpen
  \bibfield  {author} {\bibinfo {author} {\bibfnamefont {K.}~\bibnamefont {{Ta
  Phuoc}}}, \bibinfo {author} {\bibfnamefont {S.}~\bibnamefont {Corde}},
  \bibinfo {author} {\bibfnamefont {C.}~\bibnamefont {Thaury}}, \bibinfo
  {author} {\bibfnamefont {V.}~\bibnamefont {Malka}}, \bibinfo {author}
  {\bibfnamefont {A.}~\bibnamefont {Tafzi}}, \bibinfo {author} {\bibfnamefont
  {J.-P.}\ \bibnamefont {Goddet}}, \bibinfo {author} {\bibfnamefont
  {R.}~\bibnamefont {C.Shah}}, \bibinfo {author} {\bibfnamefont
  {S.}~\bibnamefont {Sebban}},\ and\ \bibinfo {author} {\bibfnamefont
  {A.}~\bibnamefont {Rousse}},\ }\bibfield  {title} {\enquote {\bibinfo {title}
  {All-optical compton gamma-ray source},}\ }\href
  {https://doi.org/10.1038/nphoton.2012.82} {\bibfield  {journal} {\bibinfo
  {journal} {Nat. Photon.}\ }\textbf {\bibinfo {volume} {6}},\ \bibinfo {pages}
  {308--311} (\bibinfo {year} {2012})}\BibitemShut {NoStop}%
\bibitem [{\citenamefont {Schroeder}\ \emph {et~al.}(2010)\citenamefont
  {Schroeder}, \citenamefont {Esarey}, \citenamefont {Geddes}, \citenamefont
  {Benedetti},\ and\ \citenamefont {Leemans}}]{schr10}%
  \BibitemOpen
  \bibfield  {author} {\bibinfo {author} {\bibfnamefont {C.~B.}\ \bibnamefont
  {Schroeder}}, \bibinfo {author} {\bibfnamefont {E.}~\bibnamefont {Esarey}},
  \bibinfo {author} {\bibfnamefont {C.~G.~R.}\ \bibnamefont {Geddes}}, \bibinfo
  {author} {\bibfnamefont {C.}~\bibnamefont {Benedetti}},\ and\ \bibinfo
  {author} {\bibfnamefont {W.~P.}\ \bibnamefont {Leemans}},\ }\bibfield
  {title} {\enquote {\bibinfo {title} {Physics considerations for laser-plasma
  linear colliders},}\ }\href {https://doi.org/10.1103/PhysRevSTAB.13.101301}
  {\bibfield  {journal} {\bibinfo  {journal} {Phys. Rev. ST Accel. Beams}\
  }\textbf {\bibinfo {volume} {13}},\ \bibinfo {pages} {101301} (\bibinfo
  {year} {2010})}\BibitemShut {NoStop}%
\bibitem [{\citenamefont {He}\ \emph {et~al.}(2016)\citenamefont {He},
  \citenamefont {Beaurepaire}, \citenamefont {Nees}, \citenamefont {Gall\'e},
  \citenamefont {Scott}, \citenamefont {P\'erez}, \citenamefont {Lagally},
  \citenamefont {Krushelnick}, \citenamefont {Thomas},\ and\ \citenamefont
  {Faure}}]{he16}%
  \BibitemOpen
  \bibfield  {author} {\bibinfo {author} {\bibfnamefont {Z.-H.}\ \bibnamefont
  {He}}, \bibinfo {author} {\bibfnamefont {B.}~\bibnamefont {Beaurepaire}},
  \bibinfo {author} {\bibfnamefont {J.~A.}\ \bibnamefont {Nees}}, \bibinfo
  {author} {\bibfnamefont {G.}~\bibnamefont {Gall\'e}}, \bibinfo {author}
  {\bibfnamefont {S.~A.}\ \bibnamefont {Scott}}, \bibinfo {author}
  {\bibfnamefont {J.~R.~S.}\ \bibnamefont {P\'erez}}, \bibinfo {author}
  {\bibfnamefont {M.~G.}\ \bibnamefont {Lagally}}, \bibinfo {author}
  {\bibfnamefont {K.}~\bibnamefont {Krushelnick}}, \bibinfo {author}
  {\bibfnamefont {A.~G.~R.}\ \bibnamefont {Thomas}},\ and\ \bibinfo {author}
  {\bibfnamefont {J.}~\bibnamefont {Faure}},\ }\bibfield  {title} {\enquote
  {\bibinfo {title} {Capturing structural dynamics in crystalline silicon using
  chirped electrons from a laser wakefield accelerator},}\ }\href
  {https://doi.org/10.1038/srep36224} {\bibfield  {journal} {\bibinfo
  {journal} {Sci. Rep.}\ }\textbf {\bibinfo {volume} {6}},\ \bibinfo {pages}
  {36224} (\bibinfo {year} {2016})}\BibitemShut {NoStop}%
\bibitem [{\citenamefont {Faure}\ \emph {et~al.}(2016)\citenamefont {Faure},
  \citenamefont {van~der Geer}, \citenamefont {Beaurepaire}, \citenamefont
  {Gall\'e}, \citenamefont {Vernier},\ and\ \citenamefont
  {Lifschitz}}]{faure16}%
  \BibitemOpen
  \bibfield  {author} {\bibinfo {author} {\bibfnamefont {J.}~\bibnamefont
  {Faure}}, \bibinfo {author} {\bibfnamefont {B.}~\bibnamefont {van~der Geer}},
  \bibinfo {author} {\bibfnamefont {B.}~\bibnamefont {Beaurepaire}}, \bibinfo
  {author} {\bibfnamefont {G.}~\bibnamefont {Gall\'e}}, \bibinfo {author}
  {\bibfnamefont {A.}~\bibnamefont {Vernier}},\ and\ \bibinfo {author}
  {\bibfnamefont {A.}~\bibnamefont {Lifschitz}},\ }\bibfield  {title} {\enquote
  {\bibinfo {title} {Concept of a laser-plasma-based electron source for
  sub-10-fs electron diffraction},}\ }\href
  {https://doi.org/10.1103/PhysRevAccelBeams.19.021302} {\bibfield  {journal}
  {\bibinfo  {journal} {Phys. Rev. Accel. Beams}\ }\textbf {\bibinfo {volume}
  {19}},\ \bibinfo {pages} {021302} (\bibinfo {year} {2016})}\BibitemShut
  {NoStop}%
\bibitem [{\citenamefont {Rigaud}\ \emph {et~al.}(2010)\citenamefont {Rigaud},
  \citenamefont {Fortunel}, \citenamefont {Vaigot}, \citenamefont {Cadio},
  \citenamefont {Martin}, \citenamefont {Lundh}, \citenamefont {Faure},
  \citenamefont {Rechatin}, \citenamefont {Malka},\ and\ \citenamefont
  {Gauduel}}]{Rigaud2010}%
  \BibitemOpen
  \bibfield  {author} {\bibinfo {author} {\bibfnamefont {O.}~\bibnamefont
  {Rigaud}}, \bibinfo {author} {\bibfnamefont {N.~O.}\ \bibnamefont
  {Fortunel}}, \bibinfo {author} {\bibfnamefont {P.}~\bibnamefont {Vaigot}},
  \bibinfo {author} {\bibfnamefont {E.}~\bibnamefont {Cadio}}, \bibinfo
  {author} {\bibfnamefont {M.~T.}\ \bibnamefont {Martin}}, \bibinfo {author}
  {\bibfnamefont {O.}~\bibnamefont {Lundh}}, \bibinfo {author} {\bibfnamefont
  {J.}~\bibnamefont {Faure}}, \bibinfo {author} {\bibfnamefont
  {C.}~\bibnamefont {Rechatin}}, \bibinfo {author} {\bibfnamefont
  {V.}~\bibnamefont {Malka}},\ and\ \bibinfo {author} {\bibfnamefont {Y.~A.}\
  \bibnamefont {Gauduel}},\ }\bibfield  {title} {\enquote {\bibinfo {title}
  {Exploring ultrashort high-energy electron-induced damage in human carcinoma
  cells},}\ }\href {https://doi.org/10.1038/cddis.2010.46} {\bibfield
  {journal} {\bibinfo  {journal} {Cell Death {\&} Disease}\ }\textbf {\bibinfo
  {volume} {1}},\ \bibinfo {pages} {e73--e73} (\bibinfo {year}
  {2010})}\BibitemShut {NoStop}%
\bibitem [{\citenamefont {Lundh}\ \emph {et~al.}(2012)\citenamefont {Lundh},
  \citenamefont {Rechatin}, \citenamefont {Faure}, \citenamefont {Ben-Ismaïl},
  \citenamefont {Lim}, \citenamefont {De~Wagter}, \citenamefont {De~Neve},\
  and\ \citenamefont {Malka}}]{lund12}%
  \BibitemOpen
  \bibfield  {author} {\bibinfo {author} {\bibfnamefont {O.}~\bibnamefont
  {Lundh}}, \bibinfo {author} {\bibfnamefont {C.}~\bibnamefont {Rechatin}},
  \bibinfo {author} {\bibfnamefont {J.}~\bibnamefont {Faure}}, \bibinfo
  {author} {\bibfnamefont {A.}~\bibnamefont {Ben-Ismaïl}}, \bibinfo {author}
  {\bibfnamefont {J.}~\bibnamefont {Lim}}, \bibinfo {author} {\bibfnamefont
  {C.}~\bibnamefont {De~Wagter}}, \bibinfo {author} {\bibfnamefont
  {W.}~\bibnamefont {De~Neve}},\ and\ \bibinfo {author} {\bibfnamefont
  {V.}~\bibnamefont {Malka}},\ }\bibfield  {title} {\enquote {\bibinfo {title}
  {Comparison of measured with calculated dose distribution from a 120-mev
  electron beam from a laser-plasma accelerator},}\ }\href
  {https://doi.org/https://doi.org/10.1118/1.4719962} {\bibfield  {journal}
  {\bibinfo  {journal} {Medical Physics}\ }\textbf {\bibinfo {volume} {39}},\
  \bibinfo {pages} {3501--3508} (\bibinfo {year} {2012})},\ \Eprint
  {https://arxiv.org/abs/https://aapm.onlinelibrary.wiley.com/doi/pdf/10.1118/1.4719962}
  {https://aapm.onlinelibrary.wiley.com/doi/pdf/10.1118/1.4719962} \BibitemShut
  {NoStop}%
\bibitem [{\citenamefont {Bulanov}\ \emph {et~al.}(1998)\citenamefont
  {Bulanov}, \citenamefont {Naumova}, \citenamefont {Pegoraro},\ and\
  \citenamefont {Sakai}}]{bula98}%
  \BibitemOpen
  \bibfield  {author} {\bibinfo {author} {\bibfnamefont {S.}~\bibnamefont
  {Bulanov}}, \bibinfo {author} {\bibfnamefont {N.}~\bibnamefont {Naumova}},
  \bibinfo {author} {\bibfnamefont {F.}~\bibnamefont {Pegoraro}},\ and\
  \bibinfo {author} {\bibfnamefont {J.}~\bibnamefont {Sakai}},\ }\bibfield
  {title} {\enquote {\bibinfo {title} {Particle injection into the wave
  acceleration phase due to nonlinear wake wave breaking},}\ }\href
  {https://doi.org/10.1103/PhysRevE.58.R5257} {\bibfield  {journal} {\bibinfo
  {journal} {Phys. Rev. E}\ }\textbf {\bibinfo {volume} {58}},\ \bibinfo
  {pages} {R5257--R5260} (\bibinfo {year} {1998})}\BibitemShut {NoStop}%
\bibitem [{\citenamefont {Tomassini}\ \emph {et~al.}(2003)\citenamefont
  {Tomassini}, \citenamefont {Galimberti}, \citenamefont {Giulietti},
  \citenamefont {Giulietti}, \citenamefont {Gizzi}, \citenamefont {Labate},\
  and\ \citenamefont {Pegoraro}}]{toma03}%
  \BibitemOpen
  \bibfield  {author} {\bibinfo {author} {\bibfnamefont {P.}~\bibnamefont
  {Tomassini}}, \bibinfo {author} {\bibfnamefont {M.}~\bibnamefont
  {Galimberti}}, \bibinfo {author} {\bibfnamefont {A.}~\bibnamefont
  {Giulietti}}, \bibinfo {author} {\bibfnamefont {D.}~\bibnamefont
  {Giulietti}}, \bibinfo {author} {\bibfnamefont {L.~A.}\ \bibnamefont
  {Gizzi}}, \bibinfo {author} {\bibfnamefont {L.}~\bibnamefont {Labate}},\ and\
  \bibinfo {author} {\bibfnamefont {F.}~\bibnamefont {Pegoraro}},\ }\bibfield
  {title} {\enquote {\bibinfo {title} {Production of high-quality electron
  beams in numerical experiments of laser wakefield acceleration with
  longitudinal wave breaking},}\ }\href
  {https://doi.org/10.1103/PhysRevSTAB.6.121301} {\bibfield  {journal}
  {\bibinfo  {journal} {Phys. Rev. ST Accel. Beams}\ }\textbf {\bibinfo
  {volume} {6}},\ \bibinfo {pages} {121301} (\bibinfo {year}
  {2003})}\BibitemShut {NoStop}%
\bibitem [{\citenamefont {Suk}\ \emph {et~al.}(2001)\citenamefont {Suk},
  \citenamefont {Barov}, \citenamefont {Rosenzweig},\ and\ \citenamefont
  {Esarey}}]{suk01}%
  \BibitemOpen
  \bibfield  {author} {\bibinfo {author} {\bibfnamefont {H.}~\bibnamefont
  {Suk}}, \bibinfo {author} {\bibfnamefont {N.}~\bibnamefont {Barov}}, \bibinfo
  {author} {\bibfnamefont {J.~B.}\ \bibnamefont {Rosenzweig}},\ and\ \bibinfo
  {author} {\bibfnamefont {E.}~\bibnamefont {Esarey}},\ }\bibfield  {title}
  {\enquote {\bibinfo {title} {Plasma electron trapping and acceleration in a
  plasma wake field using a density transition},}\ }\href
  {https://doi.org/10.1103/PhysRevLett.86.1011} {\bibfield  {journal} {\bibinfo
   {journal} {Phys. Rev. Lett.}\ }\textbf {\bibinfo {volume} {86}},\ \bibinfo
  {pages} {1011--1014} (\bibinfo {year} {2001})}\BibitemShut {NoStop}%
\bibitem [{\citenamefont {Kim}, \citenamefont {Hafz},\ and\ \citenamefont
  {Suk}(2004)}]{kim04}%
  \BibitemOpen
  \bibfield  {author} {\bibinfo {author} {\bibfnamefont {J.~U.}\ \bibnamefont
  {Kim}}, \bibinfo {author} {\bibfnamefont {N.}~\bibnamefont {Hafz}},\ and\
  \bibinfo {author} {\bibfnamefont {H.}~\bibnamefont {Suk}},\ }\bibfield
  {title} {\enquote {\bibinfo {title} {Electron trapping and acceleration
  across a parabolic plasma density profile},}\ }\href
  {https://doi.org/10.1103/PhysRevE.69.026409} {\bibfield  {journal} {\bibinfo
  {journal} {Phys. Rev. E}\ }\textbf {\bibinfo {volume} {69}},\ \bibinfo {eid}
  {026409} (\bibinfo {year} {2004})}\BibitemShut {NoStop}%
\bibitem [{\citenamefont {Chien}\ \emph {et~al.}(2005)\citenamefont {Chien},
  \citenamefont {Chang}, \citenamefont {Lee}, \citenamefont {Lin},
  \citenamefont {Wang},\ and\ \citenamefont {Chen}}]{chien05}%
  \BibitemOpen
  \bibfield  {author} {\bibinfo {author} {\bibfnamefont {T.-Y.}\ \bibnamefont
  {Chien}}, \bibinfo {author} {\bibfnamefont {C.-L.}\ \bibnamefont {Chang}},
  \bibinfo {author} {\bibfnamefont {C.-H.}\ \bibnamefont {Lee}}, \bibinfo
  {author} {\bibfnamefont {J.-Y.}\ \bibnamefont {Lin}}, \bibinfo {author}
  {\bibfnamefont {J.}~\bibnamefont {Wang}},\ and\ \bibinfo {author}
  {\bibfnamefont {S.-Y.}\ \bibnamefont {Chen}},\ }\bibfield  {title} {\enquote
  {\bibinfo {title} {Spatially localized self-injection of electrons in a
  self-modulated laser-wakefield accelerator by using a laser-induced transient
  density ramp},}\ }\href {https://doi.org/10.1103/PhysRevLett.94.115003}
  {\bibfield  {journal} {\bibinfo  {journal} {Phys. Rev. Lett.}\ }\textbf
  {\bibinfo {volume} {94}},\ \bibinfo {pages} {115003} (\bibinfo {year}
  {2005})}\BibitemShut {NoStop}%
\bibitem [{\citenamefont {Faure}\ \emph {et~al.}(2010)\citenamefont {Faure},
  \citenamefont {Rechatin}, \citenamefont {Lundh}, \citenamefont {Ammoura},\
  and\ \citenamefont {Malka}}]{faur10}%
  \BibitemOpen
  \bibfield  {author} {\bibinfo {author} {\bibfnamefont {J.}~\bibnamefont
  {Faure}}, \bibinfo {author} {\bibfnamefont {C.}~\bibnamefont {Rechatin}},
  \bibinfo {author} {\bibfnamefont {O.}~\bibnamefont {Lundh}}, \bibinfo
  {author} {\bibfnamefont {L.}~\bibnamefont {Ammoura}},\ and\ \bibinfo {author}
  {\bibfnamefont {V.}~\bibnamefont {Malka}},\ }\bibfield  {title} {\enquote
  {\bibinfo {title} {Injection and acceleration of quasimonoenergetic
  relativistic electron beams using density gradients at the edges of a plasma
  channel},}\ }\href {https://doi.org/http://dx.doi.org/10.1063/1.3469581}
  {\bibfield  {journal} {\bibinfo  {journal} {Phys. Plasmas}\ }\textbf
  {\bibinfo {volume} {17}},\ \bibinfo {eid} {083107} (\bibinfo {year}
  {2010})}\BibitemShut {NoStop}%
\bibitem [{\citenamefont {Schmid}\ \emph {et~al.}(2010)\citenamefont {Schmid},
  \citenamefont {Buck}, \citenamefont {Sears}, \citenamefont {Mikhailova},
  \citenamefont {Tautz}, \citenamefont {Herrmann}, \citenamefont {Geissler},
  \citenamefont {Krausz},\ and\ \citenamefont {Veisz}}]{schm10}%
  \BibitemOpen
  \bibfield  {author} {\bibinfo {author} {\bibfnamefont {K.}~\bibnamefont
  {Schmid}}, \bibinfo {author} {\bibfnamefont {A.}~\bibnamefont {Buck}},
  \bibinfo {author} {\bibfnamefont {C.~M.~S.}\ \bibnamefont {Sears}}, \bibinfo
  {author} {\bibfnamefont {J.~M.}\ \bibnamefont {Mikhailova}}, \bibinfo
  {author} {\bibfnamefont {R.}~\bibnamefont {Tautz}}, \bibinfo {author}
  {\bibfnamefont {D.}~\bibnamefont {Herrmann}}, \bibinfo {author}
  {\bibfnamefont {M.}~\bibnamefont {Geissler}}, \bibinfo {author}
  {\bibfnamefont {F.}~\bibnamefont {Krausz}},\ and\ \bibinfo {author}
  {\bibfnamefont {L.}~\bibnamefont {Veisz}},\ }\bibfield  {title} {\enquote
  {\bibinfo {title} {Density-transition based electron injector for laser
  driven wakefield accelerators},}\ }\href
  {https://doi.org/10.1103/PhysRevSTAB.13.091301} {\bibfield  {journal}
  {\bibinfo  {journal} {Phys. Rev. ST Accel. Beams}\ }\textbf {\bibinfo
  {volume} {13}},\ \bibinfo {pages} {091301} (\bibinfo {year}
  {2010})}\BibitemShut {NoStop}%
\bibitem [{\citenamefont {Thaury}\ \emph {et~al.}(2015)\citenamefont {Thaury},
  \citenamefont {Guillaume}, \citenamefont {Lifschitz}, \citenamefont
  {Ta~Phuoc}, \citenamefont {Hansson}, \citenamefont {Grittani}, \citenamefont
  {Gautier}, \citenamefont {Goddet}, \citenamefont {Tafzi}, \citenamefont
  {Lundh},\ and\ \citenamefont {Malka}}]{thaury15}%
  \BibitemOpen
  \bibfield  {author} {\bibinfo {author} {\bibfnamefont {C.}~\bibnamefont
  {Thaury}}, \bibinfo {author} {\bibfnamefont {E.}~\bibnamefont {Guillaume}},
  \bibinfo {author} {\bibfnamefont {A.}~\bibnamefont {Lifschitz}}, \bibinfo
  {author} {\bibfnamefont {K.}~\bibnamefont {Ta~Phuoc}}, \bibinfo {author}
  {\bibfnamefont {M.}~\bibnamefont {Hansson}}, \bibinfo {author} {\bibfnamefont
  {G.}~\bibnamefont {Grittani}}, \bibinfo {author} {\bibfnamefont
  {J.}~\bibnamefont {Gautier}}, \bibinfo {author} {\bibfnamefont {J.-P.}\
  \bibnamefont {Goddet}}, \bibinfo {author} {\bibfnamefont {A.}~\bibnamefont
  {Tafzi}}, \bibinfo {author} {\bibfnamefont {O.}~\bibnamefont {Lundh}},\ and\
  \bibinfo {author} {\bibfnamefont {V.}~\bibnamefont {Malka}},\ }\bibfield
  {title} {\enquote {\bibinfo {title} {Shock assisted ionization injection in
  laser-plasma accelerators},}\ }\href {https://doi.org/10.1038/srep16310}
  {\bibfield  {journal} {\bibinfo  {journal} {Scientific Reports}\ }\textbf
  {\bibinfo {volume} {5}},\ \bibinfo {pages} {16310} (\bibinfo {year}
  {2015})}\BibitemShut {NoStop}%
\bibitem [{\citenamefont {Swanson}\ \emph {et~al.}(2017)\citenamefont
  {Swanson}, \citenamefont {Tsai}, \citenamefont {Barber}, \citenamefont
  {Lehe}, \citenamefont {Mao}, \citenamefont {Steinke}, \citenamefont {van
  Tilborg}, \citenamefont {Nakamura}, \citenamefont {Geddes}, \citenamefont
  {Schroeder}, \citenamefont {Esarey},\ and\ \citenamefont
  {Leemans}}]{swanson17}%
  \BibitemOpen
  \bibfield  {author} {\bibinfo {author} {\bibfnamefont {K.~K.}\ \bibnamefont
  {Swanson}}, \bibinfo {author} {\bibfnamefont {H.-E.}\ \bibnamefont {Tsai}},
  \bibinfo {author} {\bibfnamefont {S.~K.}\ \bibnamefont {Barber}}, \bibinfo
  {author} {\bibfnamefont {R.}~\bibnamefont {Lehe}}, \bibinfo {author}
  {\bibfnamefont {H.-S.}\ \bibnamefont {Mao}}, \bibinfo {author} {\bibfnamefont
  {S.}~\bibnamefont {Steinke}}, \bibinfo {author} {\bibfnamefont
  {J.}~\bibnamefont {van Tilborg}}, \bibinfo {author} {\bibfnamefont
  {K.}~\bibnamefont {Nakamura}}, \bibinfo {author} {\bibfnamefont {C.~G.~R.}\
  \bibnamefont {Geddes}}, \bibinfo {author} {\bibfnamefont {C.~B.}\
  \bibnamefont {Schroeder}}, \bibinfo {author} {\bibfnamefont {E.}~\bibnamefont
  {Esarey}},\ and\ \bibinfo {author} {\bibfnamefont {W.~P.}\ \bibnamefont
  {Leemans}},\ }\bibfield  {title} {\enquote {\bibinfo {title} {Control of
  tunable, monoenergetic laser-plasma-accelerated electron beams using a
  shock-induced density downramp injector},}\ }\href
  {https://doi.org/10.1103/PhysRevAccelBeams.20.051301} {\bibfield  {journal}
  {\bibinfo  {journal} {Phys. Rev. Accel. Beams}\ }\textbf {\bibinfo {volume}
  {20}},\ \bibinfo {pages} {051301} (\bibinfo {year} {2017})}\BibitemShut
  {NoStop}%
\bibitem [{\citenamefont {Haberberger}\ \emph {et~al.}(2012)\citenamefont
  {Haberberger}, \citenamefont {Tochitsky}, \citenamefont {Fiuza},
  \citenamefont {Gong}, \citenamefont {Fonseca}, \citenamefont {Silva},
  \citenamefont {Mori},\ and\ \citenamefont {Joshi}}]{Haberberger2012}%
  \BibitemOpen
  \bibfield  {author} {\bibinfo {author} {\bibfnamefont {D.}~\bibnamefont
  {Haberberger}}, \bibinfo {author} {\bibfnamefont {S.}~\bibnamefont
  {Tochitsky}}, \bibinfo {author} {\bibfnamefont {F.}~\bibnamefont {Fiuza}},
  \bibinfo {author} {\bibfnamefont {C.}~\bibnamefont {Gong}}, \bibinfo {author}
  {\bibfnamefont {R.~A.}\ \bibnamefont {Fonseca}}, \bibinfo {author}
  {\bibfnamefont {L.~O.}\ \bibnamefont {Silva}}, \bibinfo {author}
  {\bibfnamefont {W.~B.}\ \bibnamefont {Mori}},\ and\ \bibinfo {author}
  {\bibfnamefont {C.}~\bibnamefont {Joshi}},\ }\bibfield  {title} {\enquote
  {\bibinfo {title} {Collisionless shocks in laser-produced plasma generate
  monoenergetic high-energy proton beams},}\ }\href
  {https://doi.org/10.1038/nphys2130} {\bibfield  {journal} {\bibinfo
  {journal} {Nature Physics}\ }\textbf {\bibinfo {volume} {8}},\ \bibinfo
  {pages} {95--99} (\bibinfo {year} {2012})}\BibitemShut {NoStop}%
\bibitem [{\citenamefont {Nakamura}\ \emph {et~al.}(2010)\citenamefont
  {Nakamura}, \citenamefont {Bulanov}, \citenamefont {Esirkepov},\ and\
  \citenamefont {Kando}}]{naka10}%
  \BibitemOpen
  \bibfield  {author} {\bibinfo {author} {\bibfnamefont {T.}~\bibnamefont
  {Nakamura}}, \bibinfo {author} {\bibfnamefont {S.~V.}\ \bibnamefont
  {Bulanov}}, \bibinfo {author} {\bibfnamefont {T.~Z.}\ \bibnamefont
  {Esirkepov}},\ and\ \bibinfo {author} {\bibfnamefont {M.}~\bibnamefont
  {Kando}},\ }\bibfield  {title} {\enquote {\bibinfo {title} {High-energy ions
  from near-critical density plasmas via magnetic vortex acceleration},}\
  }\href {https://doi.org/10.1103/PhysRevLett.105.135002} {\bibfield  {journal}
  {\bibinfo  {journal} {Phys. Rev. Lett.}\ }\textbf {\bibinfo {volume} {105}},\
  \bibinfo {pages} {135002} (\bibinfo {year} {2010})}\BibitemShut {NoStop}%
\bibitem [{\citenamefont {Sylla}\ \emph {et~al.}(2013)\citenamefont {Sylla},
  \citenamefont {Flacco}, \citenamefont {Kahaly}, \citenamefont {Veltcheva},
  \citenamefont {Lifschitz}, \citenamefont {Malka}, \citenamefont
  {d'Humi\`eres}, \citenamefont {Andriyash},\ and\ \citenamefont
  {Tikhonchuk}}]{sylla13}%
  \BibitemOpen
  \bibfield  {author} {\bibinfo {author} {\bibfnamefont {F.}~\bibnamefont
  {Sylla}}, \bibinfo {author} {\bibfnamefont {A.}~\bibnamefont {Flacco}},
  \bibinfo {author} {\bibfnamefont {S.}~\bibnamefont {Kahaly}}, \bibinfo
  {author} {\bibfnamefont {M.}~\bibnamefont {Veltcheva}}, \bibinfo {author}
  {\bibfnamefont {A.}~\bibnamefont {Lifschitz}}, \bibinfo {author}
  {\bibfnamefont {V.}~\bibnamefont {Malka}}, \bibinfo {author} {\bibfnamefont
  {E.}~\bibnamefont {d'Humi\`eres}}, \bibinfo {author} {\bibfnamefont
  {I.}~\bibnamefont {Andriyash}},\ and\ \bibinfo {author} {\bibfnamefont
  {V.}~\bibnamefont {Tikhonchuk}},\ }\bibfield  {title} {\enquote {\bibinfo
  {title} {Short intense laser pulse collapse in near-critical plasma},}\
  }\href {https://doi.org/10.1103/PhysRevLett.110.085001} {\bibfield  {journal}
  {\bibinfo  {journal} {Phys. Rev. Lett.}\ }\textbf {\bibinfo {volume} {110}},\
  \bibinfo {pages} {085001} (\bibinfo {year} {2013})}\BibitemShut {NoStop}%
\bibitem [{\citenamefont {Fan-Chiang}\ \emph {et~al.}(2020)\citenamefont
  {Fan-Chiang}, \citenamefont {Mao}, \citenamefont {Tsai}, \citenamefont
  {Ostermayr}, \citenamefont {Swanson}, \citenamefont {Barber}, \citenamefont
  {Steinke}, \citenamefont {van Tilborg}, \citenamefont {Geddes},\ and\
  \citenamefont {Leemans}}]{Fan-Chiang20}%
  \BibitemOpen
  \bibfield  {author} {\bibinfo {author} {\bibfnamefont {L.}~\bibnamefont
  {Fan-Chiang}}, \bibinfo {author} {\bibfnamefont {H.-S.}\ \bibnamefont {Mao}},
  \bibinfo {author} {\bibfnamefont {H.-E.}\ \bibnamefont {Tsai}}, \bibinfo
  {author} {\bibfnamefont {T.}~\bibnamefont {Ostermayr}}, \bibinfo {author}
  {\bibfnamefont {K.~K.}\ \bibnamefont {Swanson}}, \bibinfo {author}
  {\bibfnamefont {S.~K.}\ \bibnamefont {Barber}}, \bibinfo {author}
  {\bibfnamefont {S.}~\bibnamefont {Steinke}}, \bibinfo {author} {\bibfnamefont
  {J.}~\bibnamefont {van Tilborg}}, \bibinfo {author} {\bibfnamefont
  {C.~G.~R.}\ \bibnamefont {Geddes}},\ and\ \bibinfo {author} {\bibfnamefont
  {W.~P.}\ \bibnamefont {Leemans}},\ }\bibfield  {title} {\enquote {\bibinfo
  {title} {Gas density structure of supersonic flows impinged on by thin blades
  for laser–plasma accelerator targets},}\ }\href
  {https://doi.org/10.1063/5.0005888} {\bibfield  {journal} {\bibinfo
  {journal} {Physics of Fluids}\ }\textbf {\bibinfo {volume} {32}},\ \bibinfo
  {pages} {066108} (\bibinfo {year} {2020})},\ \Eprint
  {https://arxiv.org/abs/https://doi.org/10.1063/5.0005888}
  {https://doi.org/10.1063/5.0005888} \BibitemShut {NoStop}%
\bibitem [{\citenamefont {Mollica}(2016)}]{molli16}%
  \BibitemOpen
  \bibfield  {author} {\bibinfo {author} {\bibfnamefont {F.}~\bibnamefont
  {Mollica}},\ }\emph {\bibinfo {title} {Ultra-intense laser-plasma interaction
  at near-critical density for ion acceleration}},\ \href
  {http://www.theses.fr/2016SACLX058/document} {Ph.D. thesis} (\bibinfo {year}
  {2016})\BibitemShut {NoStop}%
\bibitem [{\citenamefont {Rovige}\ \emph {et~al.}(2020)\citenamefont {Rovige},
  \citenamefont {Huijts}, \citenamefont {Andriyash}, \citenamefont {Vernier},
  \citenamefont {Tomkus}, \citenamefont {Girdauskas}, \citenamefont
  {Raciukaitis}, \citenamefont {Dudutis}, \citenamefont {Stankevic},
  \citenamefont {Gecys}, \citenamefont {Ouille}, \citenamefont {Cheng},
  \citenamefont {Lopez-Martens},\ and\ \citenamefont {Faure}}]{rovige20}%
  \BibitemOpen
  \bibfield  {author} {\bibinfo {author} {\bibfnamefont {L.}~\bibnamefont
  {Rovige}}, \bibinfo {author} {\bibfnamefont {J.}~\bibnamefont {Huijts}},
  \bibinfo {author} {\bibfnamefont {I.}~\bibnamefont {Andriyash}}, \bibinfo
  {author} {\bibfnamefont {A.}~\bibnamefont {Vernier}}, \bibinfo {author}
  {\bibfnamefont {V.}~\bibnamefont {Tomkus}}, \bibinfo {author} {\bibfnamefont
  {V.}~\bibnamefont {Girdauskas}}, \bibinfo {author} {\bibfnamefont
  {G.}~\bibnamefont {Raciukaitis}}, \bibinfo {author} {\bibfnamefont
  {J.}~\bibnamefont {Dudutis}}, \bibinfo {author} {\bibfnamefont
  {V.}~\bibnamefont {Stankevic}}, \bibinfo {author} {\bibfnamefont
  {P.}~\bibnamefont {Gecys}}, \bibinfo {author} {\bibfnamefont
  {M.}~\bibnamefont {Ouille}}, \bibinfo {author} {\bibfnamefont
  {Z.}~\bibnamefont {Cheng}}, \bibinfo {author} {\bibfnamefont
  {R.}~\bibnamefont {Lopez-Martens}},\ and\ \bibinfo {author} {\bibfnamefont
  {J.}~\bibnamefont {Faure}},\ }\bibfield  {title} {\enquote {\bibinfo {title}
  {Demonstration of stable long-term operation of a kilohertz laser-plasma
  accelerator},}\ }\href {https://doi.org/10.1103/PhysRevAccelBeams.23.093401}
  {\bibfield  {journal} {\bibinfo  {journal} {Phys. Rev. Accel. Beams}\
  }\textbf {\bibinfo {volume} {23}},\ \bibinfo {pages} {093401} (\bibinfo
  {year} {2020})}\BibitemShut {NoStop}%
\bibitem [{\citenamefont {Zucker}\ and\ \citenamefont
  {Biblarz}(2002)}]{Zucker2002}%
  \BibitemOpen
  \bibfield  {author} {\bibinfo {author} {\bibfnamefont {R.~D.}\ \bibnamefont
  {Zucker}}\ and\ \bibinfo {author} {\bibfnamefont {O.}~\bibnamefont
  {Biblarz}},\ }\href@noop {} {\emph {\bibinfo {title} {Fundamentals of Gas
  Dynamics}}}\ (\bibinfo  {publisher} {John Wiley \& Sons},\ \bibinfo {year}
  {2002})\BibitemShut {NoStop}%
\bibitem [{\citenamefont {Semushin}\ and\ \citenamefont
  {Malka}(2001)}]{semu01}%
  \BibitemOpen
  \bibfield  {author} {\bibinfo {author} {\bibfnamefont {S.}~\bibnamefont
  {Semushin}}\ and\ \bibinfo {author} {\bibfnamefont {V.}~\bibnamefont
  {Malka}},\ }\bibfield  {title} {\enquote {\bibinfo {title} {High density gas
  jet nozzle design for laser target production},}\ }\href
  {https://doi.org/10.1063/1.1380393} {\bibfield  {journal} {\bibinfo
  {journal} {Rev. Sci. Instrum.}\ }\textbf {\bibinfo {volume} {72}},\ \bibinfo
  {pages} {2961--2965} (\bibinfo {year} {2001})}\BibitemShut {NoStop}%
\bibitem [{\citenamefont {Schmid}\ and\ \citenamefont
  {Veisz}(2012)}]{Schmid2012}%
  \BibitemOpen
  \bibfield  {author} {\bibinfo {author} {\bibfnamefont {K.}~\bibnamefont
  {Schmid}}\ and\ \bibinfo {author} {\bibfnamefont {L.}~\bibnamefont {Veisz}},\
  }\bibfield  {title} {\enquote {\bibinfo {title} {Supersonic gas jets for
  laser-plasma experiments},}\ }\href {https://doi.org/10.1063/1.4719915}
  {\bibfield  {journal} {\bibinfo  {journal} {Review of Scientific
  Instruments}\ }\textbf {\bibinfo {volume} {83}},\ \bibinfo {pages} {053304}
  (\bibinfo {year} {2012})},\ \Eprint
  {https://arxiv.org/abs/https://doi.org/10.1063/1.4719915}
  {https://doi.org/10.1063/1.4719915} \BibitemShut {NoStop}%
\bibitem [{\citenamefont {Liepmann}\ and\ \citenamefont
  {Roshko}(2013)}]{Liepmann2013}%
  \BibitemOpen
  \bibfield  {author} {\bibinfo {author} {\bibfnamefont {H.}~\bibnamefont
  {Liepmann}}\ and\ \bibinfo {author} {\bibfnamefont {A.}~\bibnamefont
  {Roshko}},\ }\href {https://books.google.fr/books?id=IWrCAgAAQBAJ} {\emph
  {\bibinfo {title} {Elements of Gas Dynamics}}},\ Dover Books on Aeronautical
  Engineering\ (\bibinfo  {publisher} {Dover Publications},\ \bibinfo {year}
  {2013})\BibitemShut {NoStop}%
\bibitem [{\citenamefont {Courant}\ and\ \citenamefont
  {Friedrichs}(1948)}]{courant48}%
  \BibitemOpen
  \bibfield  {author} {\bibinfo {author} {\bibfnamefont {R.}~\bibnamefont
  {Courant}}\ and\ \bibinfo {author} {\bibfnamefont {K.~O.}\ \bibnamefont
  {Friedrichs}},\ }\bibfield  {title} {\enquote {\bibinfo {title} {Supersonic
  flow and shock waves},}\ }\href@noop {} {\  (\bibinfo {year}
  {1948})}\BibitemShut {NoStop}%
\bibitem [{\citenamefont {Wilcox}\ \emph {et~al.}(1998)\citenamefont {Wilcox}
  \emph {et~al.}}]{wilcox98}%
  \BibitemOpen
  \bibfield  {author} {\bibinfo {author} {\bibfnamefont {D.~C.}\ \bibnamefont
  {Wilcox}} \emph {et~al.},\ }\href@noop {} {\emph {\bibinfo {title}
  {Turbulence modeling for CFD}}},\ Vol.~\bibinfo {volume} {2}\ (\bibinfo
  {publisher} {DCW industries La Canada, CA},\ \bibinfo {year}
  {1998})\BibitemShut {NoStop}%
\bibitem [{\citenamefont {Menter}(1994)}]{menter93}%
  \BibitemOpen
  \bibfield  {author} {\bibinfo {author} {\bibfnamefont {F.~R.}\ \bibnamefont
  {Menter}},\ }\bibfield  {title} {\enquote {\bibinfo {title} {Two-equation
  eddy-viscosity turbulence models for engineering applications},}\ }\href
  {https://doi.org/10.2514/3.12149} {\bibfield  {journal} {\bibinfo  {journal}
  {AIAA Journal}\ }\textbf {\bibinfo {volume} {32}},\ \bibinfo {pages}
  {1598--1605} (\bibinfo {year} {1994})},\ \Eprint
  {https://arxiv.org/abs/https://doi.org/10.2514/3.12149}
  {https://doi.org/10.2514/3.12149} \BibitemShut {NoStop}%
\bibitem [{\citenamefont {Primot}\ and\ \citenamefont
  {Sogno}(1995)}]{primot1995}%
  \BibitemOpen
  \bibfield  {author} {\bibinfo {author} {\bibfnamefont {J.}~\bibnamefont
  {Primot}}\ and\ \bibinfo {author} {\bibfnamefont {L.}~\bibnamefont {Sogno}},\
  }\bibfield  {title} {\enquote {\bibinfo {title} {Achromatic three-wave (or
  more) lateral shearing interferometer},}\ }\href
  {https://doi.org/10.1364/JOSAA.12.002679} {\bibfield  {journal} {\bibinfo
  {journal} {J. Opt. Soc. Am. A}\ }\textbf {\bibinfo {volume} {12}},\ \bibinfo
  {pages} {2679--2685} (\bibinfo {year} {1995})}\BibitemShut {NoStop}%
\bibitem [{\citenamefont {B{\"o}hle}\ \emph {et~al.}(2014)\citenamefont
  {B{\"o}hle}, \citenamefont {Kretschmar}, \citenamefont {Jullien},
  \citenamefont {Kovacs}, \citenamefont {Miranda}, \citenamefont {Romero},
  \citenamefont {Crespo}, \citenamefont {Morgner}, \citenamefont {Simon},
  \citenamefont {Lopez-Martens},\ and\ \citenamefont {Nagy}}]{Bohle2014}%
  \BibitemOpen
  \bibfield  {author} {\bibinfo {author} {\bibfnamefont {F.}~\bibnamefont
  {B{\"o}hle}}, \bibinfo {author} {\bibfnamefont {M.}~\bibnamefont
  {Kretschmar}}, \bibinfo {author} {\bibfnamefont {A.}~\bibnamefont {Jullien}},
  \bibinfo {author} {\bibfnamefont {M.}~\bibnamefont {Kovacs}}, \bibinfo
  {author} {\bibfnamefont {M.}~\bibnamefont {Miranda}}, \bibinfo {author}
  {\bibfnamefont {R.}~\bibnamefont {Romero}}, \bibinfo {author} {\bibfnamefont
  {H.}~\bibnamefont {Crespo}}, \bibinfo {author} {\bibfnamefont
  {U.}~\bibnamefont {Morgner}}, \bibinfo {author} {\bibfnamefont
  {P.}~\bibnamefont {Simon}}, \bibinfo {author} {\bibfnamefont
  {R.}~\bibnamefont {Lopez-Martens}},\ and\ \bibinfo {author} {\bibfnamefont
  {T.}~\bibnamefont {Nagy}},\ }\bibfield  {title} {\enquote {\bibinfo {title}
  {Compression of {CEP}-stable multi-{mJ} laser pulses down to
  4{\hspace{0.167em}}fs in long hollow fibers},}\ }\href
  {https://doi.org/10.1088/1612-2011/11/9/095401} {\bibfield  {journal}
  {\bibinfo  {journal} {Laser Physics Letters}\ }\textbf {\bibinfo {volume}
  {11}},\ \bibinfo {pages} {095401} (\bibinfo {year} {2014})}\BibitemShut
  {NoStop}%
\bibitem [{\citenamefont {Marcinkevi\v{c}ius}\ \emph
  {et~al.}(2001)\citenamefont {Marcinkevi\v{c}ius}, \citenamefont {Juodkazis},
  \citenamefont {Watanabe}, \citenamefont {Miwa}, \citenamefont {Matsuo},
  \citenamefont {Misawa},\ and\ \citenamefont {Nishii}}]{Marcinkevicius2001}%
  \BibitemOpen
  \bibfield  {author} {\bibinfo {author} {\bibfnamefont {A.}~\bibnamefont
  {Marcinkevi\v{c}ius}}, \bibinfo {author} {\bibfnamefont {S.}~\bibnamefont
  {Juodkazis}}, \bibinfo {author} {\bibfnamefont {M.}~\bibnamefont {Watanabe}},
  \bibinfo {author} {\bibfnamefont {M.}~\bibnamefont {Miwa}}, \bibinfo {author}
  {\bibfnamefont {S.}~\bibnamefont {Matsuo}}, \bibinfo {author} {\bibfnamefont
  {H.}~\bibnamefont {Misawa}},\ and\ \bibinfo {author} {\bibfnamefont
  {J.}~\bibnamefont {Nishii}},\ }\bibfield  {title} {\enquote {\bibinfo {title}
  {Femtosecond laser-assisted three-dimensional microfabrication in silica},}\
  }\href {https://doi.org/10.1364/OL.26.000277} {\bibfield  {journal} {\bibinfo
   {journal} {Opt. Lett.}\ }\textbf {\bibinfo {volume} {26}},\ \bibinfo {pages}
  {277--279} (\bibinfo {year} {2001})}\BibitemShut {NoStop}%
\bibitem [{\citenamefont {Tomkus}\ \emph {et~al.}(2018)\citenamefont {Tomkus},
  \citenamefont {Girdauskas}, \citenamefont {Dudutis}, \citenamefont
  {Ge\v{c}ys}, \citenamefont {Stankevi\v{c}},\ and\ \citenamefont
  {Ra\v{c}iukaitis}}]{Tomkus18}%
  \BibitemOpen
  \bibfield  {author} {\bibinfo {author} {\bibfnamefont {V.}~\bibnamefont
  {Tomkus}}, \bibinfo {author} {\bibfnamefont {V.}~\bibnamefont {Girdauskas}},
  \bibinfo {author} {\bibfnamefont {J.}~\bibnamefont {Dudutis}}, \bibinfo
  {author} {\bibfnamefont {P.}~\bibnamefont {Ge\v{c}ys}}, \bibinfo {author}
  {\bibfnamefont {V.}~\bibnamefont {Stankevi\v{c}}},\ and\ \bibinfo {author}
  {\bibfnamefont {G.}~\bibnamefont {Ra\v{c}iukaitis}},\ }\bibfield  {title}
  {\enquote {\bibinfo {title} {High-density gas capillary nozzles manufactured
  by hybrid 3d laser machining technique from fused silica},}\ }\href
  {https://doi.org/10.1364/OE.26.027965} {\bibfield  {journal} {\bibinfo
  {journal} {Opt. Express}\ }\textbf {\bibinfo {volume} {26}},\ \bibinfo
  {pages} {27965--27977} (\bibinfo {year} {2018})}\BibitemShut {NoStop}%
\end{thebibliography}
\end{document}